\documentclass[12pt]{article}
\usepackage{epsfig}
\usepackage[utf8]{inputenc}

\usepackage{graphicx}

\begin{document}

\title{A note on the uniqueness of the dRGT massive gravity theory } 
\author{Máximo Bañados \\
Facultad de Física\\ 
Pontificia Universidad Católica de Chile\\
Av. Vicuña Mackenna 4860, Santiago, Chile.\\
maxbanados@gmail.com}

\maketitle

\begin{abstract}
We revisit the problem of gravity coupled to a background metric $\eta_{\mu\nu}$, looking for  
ghost free interactions. It is known that elimination of the Boulware-Desewr ghost is equivalent to a certain Hessian condition on the interacting potential. We elaborate on this equation and provide a proof that dRGT potential is the unique interaction between the spacetime metric and a background field $\eta_{\mu\nu}$ which is both Lorentz invariant and ghost free. Our approach is fully non-perturbative using only the ADM formulation of general relativity. 
\end{abstract}

\section{Introduction}

It took several decades before a non-perturbative massive version of general relativity 
became available. Starting with the classic work by Fierz and Pauli back in 1936, the problem of an action for an ``interacting massive graviton" remained open for many years. Important developments where made in the seventies \cite{Boulware:1972zf,Boulware:1973my} when Boulware and Deser showed that having ghost free theory at non-linear level would not be easy. A few years before van Dam–Veltman–
Zakharov \cite{vanDam:1970vg,Zakharov:1970cc} had discovered that the $m\rightarrow 0$ limit of massive gravity would not be continuous to the $m=0$ case, adding more trouble to the theory. Back in the XXI century, a series of developments (see, for example, \cite{Creminelli:2005qk,Arkani-Hamed2003}) ended in the de Rham-Gabadadze-Tolley potential \cite{deRham:2010kj} (dRGT), the first self interacting theory for a massive spin 2 particle carrying 5 degrees of freedom, in four dimensions. Since then, massive gravity has seen many developments, and there are already excellent reviews available (see \cite{Hinterbichler2012} and \cite{deRham:2014zqa}, including complete historical and bibliographical notes). 

The derivation of the dRGT potential started with the Pauli-Fierz action, with higher order terms added order by order. This procees was `summed over' in \cite{deRham:2010kj} where a closed form, involving the square root $\eta^{\mu\nu}g_{\mu\rho} $, was presented. This form is now known as the dRGT potential. The full, non-perturbative proof, that the dRGT potential carries no ghost was worked out in 
\cite{Hassan:2011hr}, in the framework of the ADM formalism. 

In this short note we consider the converse problem. Starting from general relativity coupled to a background metric (that we choose to be $\eta_{\mu\nu}$) we demand the full theory to carry 5 degrees of freedom (or less) instead of 6. Our analysis is non-perturbative from the very beginning without assuming any particular form for the potential at the linearized level. Our motivation is to explore the possible existence of other potentials carrying five degrees of freedom, perhaps not connected perturbatively with the dRGT potential.  

The problem is mapped to a non-linear partial differential equation for the potential.  
To the best of our ability, our conclusion is that the dRGT potential is the unique Lorentz-invariant interaction between $g_{\mu\nu}$ and $\eta_{\mu\nu}$ carrying less than 6 degrees of freedom.  

\section{The problem. Equations for V($g_{\mu\nu},\eta_{\mu\nu})$}

We work on the ADM formalism. The ADM action with an interaction term reads
\begin{equation}
I = \int\,  dt\, d^n x \left( \pi^{ij} \dot g_{ij} + n {\cal H} + n^i  {\cal H}_i  - V(n,n^i,g_{ij}) \right)
\label{ADMV}
\end{equation} 
${\cal H}$ and ${\cal H}_i$ are the usual Hamiltonian constraints of general relativity. If $V$ is zero, this action carries 2 degrees of freedom. If $V$ is generic, it carries 6 degrees of freedom \cite{Boulware:1972zf,Boulware:1973my}. If $V$ is the dRGT choice, the theory carries 5 degrees of freedom. Actually to achieve a theory with 5 degrees of freedom is not difficult. The real trouble is to have, at the same time, Lorentz invariance. We elaborate this point is detail below. 
 
The proof that the action (\ref{ADMV}), for a generic potential $V$, carries 6 degrees of freedom is rather simple. One only needs to know that, if a variable inside an action can be solved { algebraically} from its own equation of motion, then one can replace its value back in the action to obtain a theory with one less variable, yet the same physics. 

Varying (\ref{ADMV}) with respecto to $N$ and $N^i$
we find
\begin{eqnarray}
{\cal H} &=&  {\partial V \over \partial n}  \\
{\cal H}_i &=&  {\partial V \over \partial n^i} \label{2}
\end{eqnarray}      
These are algebraic equations for $n,n^i$. If $V$ is a generic function, then one can solve these four equations for the four unknowns $n,n^i$ as functions of $\pi^{ij},g_{ij}$. Replacing back into the action one gets a  Hamiltonian action with a non-zero Hamiltionian for the canonical pair $\pi^{ij},g_{ij}$. Since, in four dimensions, $g_{ij}$ has 6 components this theory clearly carries 6 degrees of freedom. 

The sixth degree of freedom is problematic, as was first pointed out by Pauli and Fierz back in 1936, and fully elaborated by Boulware and Deser \cite{Boulware:1972zf,Boulware:1973my}. This fact motivated the search for a theory with 5 degrees of freedom.

One degree of freedom would be absent if there exists a change of variables $n,n^{i} \rightarrow m,m^i$ such that the Hamiltonian including the potential is linear in, say $m$, so that this field still acts as a Lagrange multiplier. (See \cite{Hassan:2011hr} for the full proof of this fact for the dRGT potential.) In that case there would be one extra constraint whose time evolution yields another constraint. Both constraints form a second class pair hence eliminating one degree of freedom.       

A more general way to express the same property is to demand that equations (\ref{2}) are not all independent and therefore do not allow for a full solution for $n,n^i$. Since the left hand side of (\ref{2}) do not depend on $n,n^{i}$, the condition implying a relation between the four equations (\ref{2}) is that the Hessian matrix of second derivatives has zero determinant 
\begin{equation}\label{hess}
\left|{\partial^2 V\over\partial n^\mu\partial n^\nu }\right| =0.
\end{equation} 
Here we are using $n^\mu = (n,n^i)$. Any potential satisfying (\ref{hess}) will give rise to a theory with less that 6 six degrees of freedom. This equation has appeared previously in the literature, for example in \cite{deRham:2014zqa,deRham:2011rn}, but, to our knowledge, its implications has not been explored in detail. So far, (\ref{hess}) has been used a a consistency check ensuring the absence of the BD mode. 

In this work we take (\ref{hess}) as the defining property (together with Lorentz invariance, discussed below) of potentials carrying less than 6 degrees of freedom.  Our goal is to solve (\ref{hess}). The dRGT potential of course satisfies (\ref{hess}). But, (\ref{hess}) is a second order non-linear equation and it is not obvious a priori that dRGT must be the unique solution. It is known that the dRGT is the most general solution connected with the Pauli-Fierz theory at linearized level. Our goal is to explore the space of solutions to (\ref{hess}) looking for theories  with 5 degrees of freedom without imposing any particular form at weak fields.  Our conclusion, however, is that it is indeed the unique solution. The details of the proof are somehow interesting and are presented in the following sections.      

\section{Lorentz symmetry}

Equation (\ref{hess}) is not the full story. In fact it is easy to find many solutions. For example, 
\begin{eqnarray}
V_1 &=& (n + a) F_1\left( {n^i + b^i \over n + a },g_{ij} \right), \label{nfn} \\
V_2 &=&  F_2(a_0 + a_\mu n^\mu,g_{ij}), \label{aa} \\ 
V_3 &=&  f_0(n^i,g_{ij}) + f_1(n^i,g_{ij})n + f_{2}(n^i,g_{ij}) n^2 + f_{3}(n^i,g_{ij}) n^3 + \cdots, \label{sols}
\end{eqnarray} 
where $F_1,F_2$ are arbitrary functions of their arguments\footnote{$a,b^i,a_0$, etc, are ``constants" - do not depend on $n^\mu$, they could depend on $g_{ij}$.} are all solutions to (\ref{hess}). The series $V_3$ is a solution after imposing some conditions among $f_{n}(n^i,g_{ij})$ following from (\ref{hess}).  

In this sense, to produce theories of gravity with 5 degrees of freedom is not difficult at all.  The real problem is to achieve 5 degrees of freedom simultaneously with the background symmetries. 

The simplest example, already not easy to implement, is to assume that the mass term is built from an interaction between the dynamical metric $g_{\mu\nu}$ and a flat background $\eta^{\mu\nu} $, keeping Lorentz invariance. This is where the problem becomes complicated. In fact, none of the solutions (\ref{nfn})-(\ref{sols}) is Lorentz invariant. (Apart from $d=2$, which stands apart. We comment on this case below).

To achieve Lorentz symmetry, $V$ must be a function of the form (in dimension $d$), 
\begin{equation}\label{V(q)}
V = V(q_1,q_2,...q_d)
\end{equation}  
where $q_i$ are traces of  the matrix $Y^{\mu}_{\ \nu} \equiv \eta^{\mu\rho}g_{\rho\nu}$. Thus, our problem is to find solutions to the equations (\ref{hess}) which, at the same time, can be written in the form (\ref{V(q)}). This turns out to be a highly complicated problem.

\section{d=3}

The problem we face is mathematically complicated and for that reason we consider the $d=3$ case. We do not expect any new effects at $d=4$. The $d=2$ case, on the other hand, is too simple. [It is easy to see that the only solution to (\ref{hess}) at $d=2$ is of the form (\ref{nfn}), and coincides with the $d=2$ dRGT potential.]  

Consider a three-dimensional $g_{\mu\nu}$ metric written in ADM form,
\begin{equation}\label{ADM}
ds^2  = -n^2 dt^2 + g_{ij}(dx^i + n^i)(dx^j + n^j dt). 
\end{equation} 
The 2-dimensional metric $g_{ij}$ has three independent components $g_{11},g_{12},g_{22}$ and it is convenient to write them in terms of the trace $(x)$, determinant $(y)$, and an extra variable $z$ via:
\begin{eqnarray}\label{xyz}
g_{11} = x+z, \ \ \   g_{22} = x-z, \ \ \ \  g_{12} = \sqrt{-y + x^2 - z^2}
\end{eqnarray} 
Our first task in to calculate the invariants of 
\begin{equation}\label{Y}
Y^{\mu}_{\ \nu} \equiv \eta^{\mu\rho} g_{\rho \nu}
\end{equation} 
(later we also introduce the invariants of $\sqrt{Y}$).  In three dimensions $Y$ has only three independent traces. 
The standard choice is\footnote{These combinations are useful because, among other things, if $Y$ is diagonalized with eigenvalues $\lambda_1,\lambda_2,\lambda_3$ the invariants $q_1,q_2,q_3$ are the symmetric polynomials,   
$$
q_1 =  \lambda_1 + \lambda_2 + \lambda_3, \ \  q_2 =  \lambda_1 \lambda_2 + \lambda_1\lambda_3 + \lambda_2\lambda_3,  \ \ 
q_3 =  \lambda_1\lambda_2\lambda_3.$$ }
\begin{eqnarray}
q_1 &=& \mbox{Tr} (Y) \nonumber\\
q_2 &=& -{1 \over 2}\mbox{Tr} (Y^2)+ {1 \over 2} \mbox{Tr}(Y)^2 \nonumber \\
q_3 &=& {1 \over 3}\mbox{Tr} (Y^3)  - {1 \over 2}\mbox{Tr}(Y)\mbox{Tr}(Y^2) + {1 \over 6} \mbox{Tr}(Y)^3  \label{q}
\end{eqnarray} 
Directly from (\ref{ADM}) and (\ref{q}) we find the explicit values for the invariants
\begin{eqnarray}
q_1 &=&  n^2-x(n_1^2 + n_2^2) -z(n_1^2-n_2^2) -2n_1n_2\,\sqrt {-{y}^{2}+{x}^{2}-{z}^{2}}+2x \nonumber\\
q_2 &=&  y^2(1-n_2^2- n_1^2)+2 n^2x \nonumber\\
q_3 &=&   n^2 y^2 \label{qadm}
\end{eqnarray}   
where we recall the redefinitions (\ref{xyz}). 

The problem is now direct. We consider functions $V(q_1,q_2,q_3)$, where $q_1,q_2,q_3$ depend on the lapse $n$ and shift $n^i$ as in (\ref{qadm}). We our goal is to find the solutions $V(q_1,q_2,q_3)$ of (\ref{hess}).

In practice, it is easier to do a coordinate change taking $q_1,q_2,q_3$ as the independent coordinates, using (\ref{qadm}) to express $n,n_1,n_2$ in terms of $q_1,q_2,q_3$ and write
\begin{eqnarray}
{\cal E} &\equiv & \left|{\partial^2 V(n^\mu) \over \partial n^\mu\partial n^\nu}\right| \nonumber\\
 &=& \left|{\partial q_i \over \partial n^\mu}{\partial q_j \over \partial n^\nu}{\partial^2 V(q_i)\over \partial q_i\partial q_j}  + {\partial^2 q_i \over \partial n^\mu \partial n^\nu} {\partial V(q_i) \over \partial q_i} \right| =0 \label{hessq}
\end{eqnarray} 
In this equation, the Jacobians ${\partial q_i \over \partial n^\mu}$ must be written in terms of $q_i$ using (\ref{q}). (Note that the inverse involves a quadratic equation. We have checked that both branches give similar results). A complication arises, however, because $n^\mu$, as a function of $q_i$, contains also a dependence on the metric components $x,y,z$. This means that equation (\ref{hessq}) cannot be written purely in terms of $q_1,q_2,q_3$ but it contains a dependence on $x,y,z$ as well.  

Indeed, (\ref{hessq}) has the form, 
\begin{eqnarray}
{\cal E} &=& {\cal E}_{1}(V) + {\cal E}_{2}(V)\,x^2 + {\cal E}_{3}(V)\,x^4 + {\cal E}_{4}(V)\,x^6 \nonumber\\
         &+& \left( {\cal E}_{5}(V) + {\cal E}_{6}(V)\,x^2 + {\cal E}_{7}(V)\,x^4\right)y \nonumber\\ 
         &+& \left({\cal E}_{8}(V) + {\cal E}_{9}(V) x^2\right)\,y^2 + {\cal E}_{10}(V)\,y^3 =0 \label{ee0}
\end{eqnarray} 
where all ${\cal E}_{n}(V)$ depend on the first and second derivatives of $V(q_i)$ with respect to $q_i$,   plus the variables $q_i$ themselves but not $x,y,z$. (see Appendix \ref{equations} for the explicit expressions). Note that $z$ has dropped out. Since $V(q_i)$ cannot depend on $x,y$,  all ten coefficients 
${\cal E}_{n}(V)$ in (\ref{ee0}) must vanish separately. Therefore, a solution to our problem will be a function $V(q_i)$ satisfying ten equations 
\begin{equation}
{\cal E}_{n}(V)=0, \ \ \ \ \ \  n=1,2,3,...,10 \label{ee}
\end{equation} 
We conclude that Lorentz invariance severely restricts the ghost free potentials. We have gone from one equation, (\ref{hess}), to a set of 10 non-linear differential equations! It is remarkable that a solution exists at all. 

The system of equations (\ref{ee}) -- ten equations for a single function $V(q)$ --looks like a horribly over determined problem. However, we know that there exists at least one solution to the problem, the dRGT potential. Thus, there must be relations between these equations. 

In the following we shall prove that the dRGT potential does indeed solve all equations, and secondly, we prove that there are no other solutions, at least under some reasonable assumptions that we display below.

\section{Invariants of $\sqrt{\eta^{-1}g }$ and the dRGT solution.}

Before proceeding, it useful to consider the above equations in a different coordinate system. We have chosen to implement Lorentz invariance by taking the potential as a function of the invariant of $Y = \eta^{-1}g $. This is useful because these invariants depend on a simple way in the metric $g_{\mu\nu}$, as shown in (\ref{q}). 

Nevertheless, it is also useful to consider the invariants of
\begin{equation}
X = \sqrt{\eta^{-1}g }. 
\end{equation}  
because the dRGT potential depends easily on them. Just as we did for the invariants of $Y = \eta^{-1}g $, 
we define,
\begin{eqnarray}
k_1 &=&  \mbox{Tr}(X)  \nonumber\\
k_{2} &=& -{1 \over 2} \mbox{Tr}(X^2) + {1 \over 2} (\mbox{Tr}\, X)^2 \\
k_3 &=&  {1 \over 3} \mbox{Tr}(X^3) + {1 \over 6} (\mbox{Tr}\, X)^3   - {1 \over 2} \mbox{Tr(X)}\mbox{Tr}(X^2) 
\end{eqnarray}   
The dRGT potential is simply a linear combination\footnote{There are two changes here with respect to the usual literature. We have considered the square root of $\eta^{-1}g $ instead of $g^{-1}\eta.$ Second, the potential (\ref{VdRGT}) does not include a $\sqrt{-g}$ in front. These two changes cancel each other.}
\begin{equation}\label{VdRGT}
V_{\mbox{dRGT}}(k_i) = a_1 k_1 + a_2 k_2 + a_3 k_3
\end{equation} 
where $a_1,a_2,a_3$ are constants with the appropriated dimensions. The invariants $k_i$ are more difficult to compute, as functions of the metric components, because they involve the square root of a matrix.  
The good news is that there is an extremely simple relation between $q_1,q_2,q_3$ and $k_1,k_2,k_3$, namely, 
\begin{eqnarray}
q_1 &=& k_1^2 - 2k_2 \nonumber\\
q_2 &=& k_2^2 - 2k_3 k_1 \nonumber\\
q_3 &=& k_3^2 \label{qk}
\end{eqnarray} 
(These relations are easily obtained going to the diagonal frame.) 

Using the ``coordinate change" (\ref{qk}) we can expresse all 10 equations ${\cal E}_{n}$ in the new coordinates. The equations in the $k$-system are even longer than those in the $q$-system, which are already quite long (see appendix),  so we do not include them here explicitly. We mention that they have the structure, 
\begin{eqnarray}\label{eqk}
{\cal E}'_{n} =  \sum {\partial^2 V \over \partial k_{i_1}\partial k_{i_2} }{\partial^2 V \over \partial k_{i_3} \partial k_{i_4} }\left ( a_{n,i_1 i_2i_3i_4i_5i_6} {\partial^2 V \over \partial k_{i_5} \partial k_{i_6} }  + b_{n,i_1 i_2i_3i_4i_5}{\partial V \over \partial k_{i_5} } \right) = 0 \ 
\end{eqnarray} 
Since (\ref{VdRGT}) is linear in the $k_i$, all second derivatives vanish and the equation (\ref{eqk}) is solved term by term. 

Observe that (\ref{eqk}) are second other differential equations and one could expect 6 integration constants, and not three, as in (\ref{VdRGT}). On the other hand, (\ref{eqk}) are 10 equations (!), so it is remarkable that there exists a solution at all. In any case, there is no reason yet to claim that the linear solution (\ref{VdRGT}) is the most the most general solution. We now prove this statement.   

A first hint of uniqueness comes from the linearization of (\ref{eqk}) around the dRGT solution. Consider 
\begin{equation}
V(k_{1},k_{2},k_{3}) =  a_1 k_1 + a_2 k_2 + a_3 k_3 + \epsilon\, f(k_1,k_2,k_3)
\end{equation} 
where $\epsilon$ is a small parameter. Plugging this into (\ref{eqk}) and keeping linear terms in $\epsilon$ one finds six equations which, by taking appropriated linear combinations, become, 
\begin{equation}
\partial_{11}f =0 , \ \ \partial_{12}f =0 , \ \ \partial_{13}f =0 , \ \ \partial_{22}f =0 , \ \ \partial_{23}f =0 , \ \ \partial_{33}f =0 \ \ 
\end{equation}  
(here $\partial_{ij}f \equiv {\partial^2 f \over \partial k_i \partial k_j}$). Thus, all second derivatives vanish and  we conclude that the only linear  fluctuation of (\ref{VdRGT}) is a modification of the integration constants $a_1,a_2,a_3$.


\section{Series solutions, a subtle effect}

Linearization does not necessarily exhaust all solutions. In this section we explore series solutions to (\ref{hessq}), confirming that the only solution to all equations is (\ref{VdRGT}).

We have written the equations in two coordinate systems. The system $q_1,q_2,q_3$ associated to the invariants of $Y=\eta^{-1}g $, and the system $k_1,k_2,k_3$ associated to the invariants of $X=\sqrt{\eta^{-1}g} $. Both systems are related by (\ref{qk}). The equations in the system $q_i$ are written in the appendix \ref{equations}. The corresponding equations in the $k_i$ are more complicated; their structure is displayed in (\ref{eqk}). 
 
The dRGT potential (\ref{VdRGT}) has a very simple expression in the $k_i$ system and solves (\ref{eqk}). 
It is natural then to look for series solutions in this system to have a direct comparison with the dRGT solution.
It turns out that the interpretation of the series in the $k_i$ system is rather curious and initially confusing. To double check our results we have also studied the series in the $q_i$ system in Sec. \ref{seq}, with the same conclusions. 

\subsection{Quadratic and Linear equations}

Before starting the analysis of Taylor series solutions, one must make sure that the equations accept such solutions at  the chosen point. Our first task is to find a point $q^{(0)}_i$ which does not have a Fucsian-like singularity making sure that the equations accepts a Taylor expansion. 

To avoid this problem, we could expand around a generic point $q_1,q_2,q_3$.  However, given the length of equations, the problem very quickly becomes out of control. The point $q_1=0,q_2=0,q_3=0$, on the other hand is preferable, but it may be a non-trivial point not accepting a Taylor expansion. 

Our strategy is the following. Taking linear combinations (with non-constant non-zero coefficients) of the equations displayed in Appendix \ref{equations} we can produce 6 quadratic equations which are simpler. These are displayed in Appendix \ref{Q}. Then, taking linear combinations of the quadratic equations we can produce 2 linear equations,
\begin{eqnarray}
{\cal L}_1 &=&  -V_{1}-4V_{13}q_{3}-4q_{2}V_{12}-2q_{3}V_{22}-2q_{1}V_{11} =0  \label{L1}\\
{\cal L}_2 &=&  -V_{2}+2V_{11}-4q_{3}V_{23}-2q_{2}V_{22}  =0, \label{L2}
\end{eqnarray} 
where $V_{ij} = {\partial V(k) \over \partial q_i \partial q_j}$. The combinations leading to (\ref{L1}) and (\ref{L2}) are quite involved, we do not include the details here. Using (\ref{qk}) one can easily write (\ref{L1}),(\ref{L1}) in the system $k_i$. 

The linear equations should not be miss understood. They do not generate the whole series of original equations displayed in Appendix \ref{equations}). If all equations (\ref{wq}) are solved, then the linear equations are also solved. But the converse is not true. The linear equations may accept solutions which do not solve the whole system.  However, since all solutions must solve the linear ones, we can study the structure of singularities, at a given point, looking at those equations.  

Since (\ref{L1}) and (\ref{L2}) are second order differential equations we expect, for each equation, a series where two coefficients are left free, and all other determined. For example, expanding around $k_1=0$, we present a series 
\begin{equation}\label{s1}
V(k_1,k_2,k_3) = V_0(k_2,k_3) +  V_1(k_2,k_3)k_1 +  V_2(k_2,k_3)k_1^2 +  V_3(k_2,k_3)k_1^3  + \cdots  
\end{equation} 
If the equations accept a Taylor expansion around the point $k_1=0$, then plugging (\ref{s1}) into (\ref{L1}) and (\ref{L2}) (independently) should generate a series where two of the functions $V_i(q_2,q_3)$ are left free, and all others are determined. If this is case, the point $k_1=0$ is a regular one. 

We have checked that this is indeed the case for the points $k_1=0$ and $k_2=0$, but not for $k_3=0$. The coordinate $k_3$ is different. Plugging a series of the form (\ref{s1}) interchanging $q_1 \leftrightarrow q_3$ one finds conditions over all coefficients. In conclusion, the equations do not accept a Taylor series around $k_3=0$. This should not be too surprising, $k_3$ is the (square root) determinant of $g_{\mu\nu}$! For our purposes here we simply do a shift and expand around $k_3=1$\footnote{To stress the point made in the text, we could have considered a Taylor series $V(k_1,k_2,k_3) = \sum_{{n_1,n_2,n_3}} f_{n_1,n_2,n_3} (k_1-c_1)^{n_1} \, (k_2-c_2)^{n_2} \, (k_3-c_3)^{n_3}$ with arbitrary $c_i$, avoiding the problem of Fuchsian singularities. However, given the length of the equations, the system gets out of control with arbitrary values for all $c_i$. As explained in the text, the choice $c_1=c_2=0$ and  $c_3=1$ does give a Taylor series and the equations are tractable.   } . 

Everything we have said for the $k_i$ system is also true for the $q_i$ system.

\subsection{Linear, plus an error}
\label{sek}

Having established that the point $k_1=0,k_2=0,k_3=0$ is a regular point (from now on we have made the shift $k_3 \rightarrow k_3 +1$, without changing the name, everywhere and expand around $k_3=0$) accepting a Taylor series 
 we set the ansatz
\begin{equation}
V(k_1,k_2,k_3) = \sum_{{n_1,n_2,n_3} \in N} f_{n_1,n_2,n_3} k_1^{n_1} \, k_2^{n_2} \, k_3^{n_3}
\end{equation} 
We plug this expansion in all equations and order by order we determine the coefficients $f_{n_1,n_2,n_3}$. 
After considerable (symbolic computational) work we have found a solution up to order 6. All higher order coefficients can be expressed in terms of the lower ones, that we have rename, 
$$
f_{1,0,0} = a_1, \ \ \  f_{0,1,0} = a_2, \ \ \  f_{0,0,1} = a_3.  
$$
The series, to order 3 is, 
\begin{eqnarray}
V&=&V_0 + { a_1}\,{ k_1}+{ a_2}\,{ k_2}+{ k_3}\,{ a_3}
\nonumber\\ 
&-&{\frac {41}{15}}\,{ a_1}
\,{{ k_1}}^{2} -{\frac {41}{15}}\,{ a_2}\,{{ k_2}}^{2}+{\frac {34
}{15}}\,{ a_1}\,{{ k_2}}^{2}+{1 \over 3}\,{{ k_3}}^{2}{ a_1}-{2 \over 3}\,{{
 k_3}}^{2}{ a_2}-{\frac {7}{15}}\,{ a_1}\,{{ k_2}}^{3}
\nonumber\\
&+&{\frac {8}{15}}\,{ a_2}\,{{ k_2}}^{3}
  -1/2\,{{ k_3}}^{3}{ a_3}+{\frac 
{19}{15}}\,{{ k_3}}^{3}{ a_1}-{\frac {26}{15}}\,{{ k_3}}^{3}{
 a_2}-{2 \over15}\,{ a_1}\,{ k_1}\,{ k_2}
\nonumber\\
&+&{\frac {4}{15}}\,{ a_2}\,
{{ k_1}}^{2}{ k_2}-{2 \over15}\,{ a_2}\,{ k_1}\,{ k_2}-{\frac {7}{
30}}\,{ a_1}\,{{ k_1}}^{2}{ k_2} -{8 \over 5}\,{ k_3}\,{ a_1}\,{ 
k_1}-{1 \over 15}\,{ k_3}\,{ a_2}\,{{ k_1}}^{2}
\nonumber\\
&+&{\frac {12}{5}}\,{ k_3}
\,{ a_2}\,{ k_1}-{\frac {7}{30}}\,{ a_2}\,{{ k_2}}^{2}{ k_1}
+ {\frac {4}{15}}\,{ a_1}\,{{ k_2}}^{2}{ k_1}+{\frac {23}{30}}\,{
{ k_3}}^{2}{ a_1}\,{ k_1} -{\frac {11}{15}}\,{{ k_3}}^{2}{ 
a_2}\,{ k_1}
\nonumber\\
&-&{\frac {14}{5}}\,{ k_3}\,{ a_1}\,{ k_2}+{\frac {16
}{5}}\,{ k_3}\,{ a_2}\,{ k_2} +{\frac {7}{30}}\,{ k_3}\,{ a_1
}\,{{ k_2}}^{2}-{\frac {4}{15}}\,{ k_3}\,{ a_2}\,{{ k_2}}^{2}-
{\frac {14}{15}}\,{{ k_3}}^{2}{ a_1}\,{ k_2}
\nonumber\\
&+&{\frac {16}{15}}\,{{ k_3}}^{2}{ a_2}\,{ k_2}-{1 \over 15}\,{ k_3}\,{ a_1}\,{{ k_1}}^{
2}+{\frac {14}{15}}\,{ k_3}\,{ a_1}\,{ k_1}\,{ k_2}-{
\frac {16}{15}}\,{ k_3}\,{ a_2}\,{ k_1}\,{ k_2} \  
\nonumber\\
&-&{\frac {7}{15
}}\,{ a_2}\,{{ k_1}}^{3}+{\frac {34}{15}}\,{ a_2}\,{{ k_1}}^{2
}+{\frac {8}{15}}\,{ a_1}\,{{ k_1}}^{3} + {\cal O}(k)^{4}
\label{Vs}
\end{eqnarray}   
The first line (linear) coincides exactly with the dRGT potential. 
This seems to indicate that (\ref{Vs}) is a different solution to the problem which, for small values of $k_i$, goes back to dRGT.  This is however not the right interpretation to (\ref{Vs}), as we now explain. 

One should suspect that (\ref{Vs}) is hiding something because dRGT is itself an exact solution to the whole problem.  Therefore, if (\ref{Vs}) was a new solution there should be a free parameter enabling to set to zero all high order terms and going back to dRGT. But this is not possible because all higher order terms in (\ref{Vs}) depend only on $a_1,a_2,a_3$.  

What happens is that all higher order terms in (\ref{Vs}) are a pure artefact of the Taylor expansion. The solution (\ref{Vs}) is simply dRGT plus an error. This can be seen, to this order, by noticing that (\ref{Vs}) can be factorized as  
\begin{eqnarray}
V&=&{ V_0}+{2 \over 5 }\,{ a_1}+{2 \over 5 }\,{ a_2}+ \left( {\frac {22}{5}}\,{ a_2}
-{\frac {23}{5}}\,{ a_1} \right)  \left( { k_1}-2 \right) 
\nonumber\\
&+&  \left( -{\frac {23}{5}}\,{ a_2}+{\frac {22}{5}}\,{ a_1} \right) 
 \left( { k_2}-2 \right) + \left( { a_3}-{\frac {22}{5}}\,{ a_1} {\frac {28}{5}}\,{ a_2} \right) { k_3}
\nonumber\\
&+&  
 \left( -{\frac {7}{30}}\,{ a_1}+{\frac {4}{15}}\,{ a_2} \right)
 \left( { k_2}-2 \right) \left( { k_1}-2 \right) ^{2}+ \left( -{1 \over 15 }\,{ a_1}-{1 \over 15 }\,{ a_2}
 \right)  \left( { k_1}-2 \right) ^{2}{ k_3} \nonumber\\
&+&  \left( -{\frac {7}{30}}\,{ a_2}+{\frac {4}{15}}\,{ a_1} \right)  \left( { k_1}-2
 \right)  \left( { k_2}-2 \right) ^{2}+ \left( {\frac {23}{30}}\,{
 a_1}- {\frac {11}{15}}\,{ a_2} \right)  \left( { k_1}-2 \right) 
{{ k_3}}^{2}
\nonumber\\ 
& +& 
 \left( {\frac {7}{30}}\,{ a_1}-{\frac {4}{15}}\,{
 a_2} \right) { k_3}\, \left( { k_2}-2 \right) ^{2} + \left( -{
\frac {14}{15}}\,{ a_1}+{\frac {16}{15}}\,{ a_2} \right) {{ k_3}
}^{2} \left( { k_2}-2 \right) 
\nonumber\\
&+&
 \left( -{\frac {7}{15}}\,{ a_2}+{
\frac {8}{15}}\,{ a_1} \right)  \left( { k_1}-2 \right) ^{3} 
 \left( -{\frac {7}{15}}\,{ a_1}+{\frac {8}{15}}\,{ a_2} \right) 
 \left( { k_2}-2 \right) ^{3}
\nonumber\\ 
&+& \left( {\frac {14}{15}}\,{ a_1}-{
\frac {16}{15}}\,{ a_2} \right) { k_3}\, \left( { k_1}
 - 2
 \right)  \left( { k_2}-2 \right) + \left( -{1 \over 2}\,{ a_3}+{\frac {19
}{15}}\,{ a_1}-{\frac {26}{15}}\,{ a_2} \right) {{ k_3}}^{3} \nonumber\\
\label{Vsf}
\end{eqnarray} 
The first two lines are the linear terms in $k_i$ plus a constant redefinition of $V_0$. Up to redefinitions, these two lines represent the dRGT solution.  

All other terms have the form $(k_1-2)^{n_1}(k_2-2)^{n_2}k_3^{n_3} $ with $n_1+n_2+n_3=3$. This could be seen as a coincidence at this order, but it is not. When expanded to order 4, the solution again can be written as a {\it linear}  piece plus terms $(k_1-2)^{n_1}(k_2-2)^{n_2}k_3^{n_3} $ with $n_1+n_2+n_3=4$. We have checked that the same effect happens again at order 5 and 6, and one can presume (see next section) that it continues at higher orders. 

In conclusion, if we interpret (\ref{Vsf}) as a series around the point $k_1=2,k_2=2,k_3=0$, this is simply the dRGT linear solution, plus an error made in the Taylor expansion.  A Taylor series can indeed be deceptive.


\section{Series in the $q_i$ system}
\label{seq}

We have studied series expansions for the equations determining the ghost free potentials and found a curios mathematical effect concerning Taylor series. After expanding around the point $k_1=0,k_2=0,k_3=0$ and obtaining a series with high order corrections (beyond the linear piece), we saw that these new terms could be seen as a pure error, when interpreting the series around a different point $k_1=2,k_2=2,k_3=0$.  

To double check the correctness of this interpretation we have studied series expansions in the $q_i$ system, arriving at the same conclusion.  

Making sure that the equations have a regular point at $q_1=0,q_2=0,q_3=1$, in the same way as we did in the previous paragraph, we set a series of the form
\begin{equation}
V(q_1,q_2,q_3) = \sum_{{n_1,n_2,n_3} \in N} f_{n_1,n_2,n_3} q_1^{n_1} \, q_2^{n_2} \, (q_3-1)^{n_3}
\end{equation} 
and plugging in the equations (\ref{wq}) we proceed order by order to solve for the coefficients $f_{n_1,n_2,n_3}$. We have proceed again to order 6. Here we display the form of the solution to order 3,  
\begin{eqnarray}
V(q_i) &=& V_0 +{b_1}\,{ q_1}+{ b_2}\,{ q_2}+{ b_3}\, \left( { 
q_3}-1 \right)
\nonumber\\
 &+&
 \left( -{1 \over 9}\,{ b_1}-{1 \over 9}\,{ b_2} \right) { q_1}\,
{ q_2}+ \left( {1 \over 9}\,{ b_2}-{2 \over 9}\,{ b_1} \right) { q_1}\,
 \left( { q_3}-1 \right)  
\nonumber\\ 
  &+&
 \left( -{1 \over 36}\,{ b_2}-{1 \over 9}\,{ b_1} \right) 
{{ q_1}}^{2}+ \left( -{1 \over 9}\,{ b_2}-{1 \over 9}\,{ b_1}-{1 \over 4}\,{ b_3}
 \right)  \left( { q_3}-1 \right) ^{2}
\nonumber\\
 &+& 
 \left( -{1 \over 36}\,{ b_1}-{1 \over 9}\,{ b_2}
 \right) {{ q_2}}^{2}+ \left( -{\frac {5}{18}}\,{
 b_2}-{1 \over 9}\,{ b_1} \right) { q_2}\, \left( { q_3}-1 \right) 
 \nonumber\\ 
 &+&
 \left( {\frac {7}{54}}\,{ b_1}-{\frac {7}{108}}\,{ b_2} \right) {
 q_1}\, \left( { q_3}-1 \right) ^{2}+ \left( {1 \over 36}\,{ b_1}+{
\frac {1}{108}}\,{ b_2} \right) {{ q_1}}^{3} 
\nonumber\\
 &+&
\left( {\frac {11}{108}}\,{ b_2}+{\frac {5}{108}}\,{ b_1} \right) + {{ q_2}}^{2}
 \left( { q_3}-1 \right) + \left( {\frac {37}{216}}\,{ b_2}+{
\frac {11}{108}}\,{ b_1} \right) { q_2}\, \left( { q_3}-1
 \right) ^{2}
\nonumber\\
 &+&
 \left( {\frac {7}{108}}\,{ b_1}-{\frac {1}{216}}\,{
 b_2} \right) {{ q_1}}^{2} \left( { q_3}-1 \right) + \left( {1 \over 27}
\,{ b_1}+{\frac {11}{216}}\,{ b_2} \right) { q_1}\,{{ q_2}}^{2}
\nonumber\\
 &+& 
 \left( {1 \over 27}\,{ b_2}+{\frac {11}{216}}\,{ b_1} \right) {{ q_1}
}^{2}{ q_2}+ \left( {1 \over 18}\,{ b_2}+{1 \over 9}\,{ b_1} \right) { q_1}\,{q_2}\, 
\left( { q_3}-1 \right) 
\nonumber\\ 
 &+&
\left( {\frac {1}{108}}\,{ b_1}+{1 \over 36}\,{ b_2} \right) {{ q_2}}^{3}  + \left( {\frac {2}{27}}\,{b_2}+{\frac {5}{54}}\,{ b_1}+{1 \over 8}\,{ b_3} \right)  
\left( { q_3}-1 \right)^{3}
\label{sq}
\end{eqnarray} 

The main question we need to ask is whether this function is different or not from the dRGT solution (\ref{VdRGT}). 
In the coordinates $q_i$, the dRGT is not linear! The comparison is more complicated.  

First of all, we have checked  that this solution is exactly the same series that we found in the previous paragraph in the $k_i$-system. That is, using (\ref{qk}) we can write (\ref{sq}) in terms of $k_i$ and comparing with (\ref{Vs}) we find exactly the same series. 

Now, to compare with the dRGT directly using (\ref{sq}) we invert the relation between the coordinate systems (\ref{qk}) expressing the $k_i$'s as functions of the $q_i$'s (gives a fourth degree equation that can be managed).   This gives $k_i(q_j)$ and we can build 
the dRGT solution in the $q_j$ system
\begin{equation}\label{dr}
V_{\mbox{dRGT}}(q_j) = A_1\, k_1(q_j) + A_2\, k_2(q_j) + A_3\, k_3(q_j) 
\end{equation} 
Now expand around $q_1=0,q_2=0$ and $q_3=1$ to find exactly the series (\ref{sq}), after identifying the coefficients $b_1,b_2,b_3$ appearing in (\ref{sq}) with linear (invertible) combinations of $A_1,A_2,A_3$ appearing in (\ref{dr}).

\section{Conclusions}

In this work we have explored whether the de Rham-Gabadazde-Tolley (dRGT) theory is the only solution to a ghost free massive gravity. The perturbative answer to this question was known to be affirmative. 
Our approach is totally non-perturbative and rely only on the equation (\ref{hess}) and its solutions, without assuming any a priori behaviour near the background. To the best of our ability, we have reach the same conclusion: the dRGT is the only ghost free potential.   

Our results are complimentary with the analysis presented in \cite{Hassan:2011hr}. 
In that reference the dRGT is assumed and a full non-perturbative proof of the absence of the ghost is presented. We do the converse statement: We assume the absence of the ghost, that is assume that the potential satisfies (\ref{hess}), and prove that the only solution is the dRGT potential. 

We have only considered the $d=3$ case, which is already quite involved mathematically. At $d=4$, the most difficult part will be to find the linear combinations yielding the quadratic and then linear equations, displayed in Appendices \ref{Q} and \ref{L}. This step is important to check the validity of the Taylor expressions. We shall leave this for the future. In any case, the $d=4$ case should have no surprises other than longer expressions.

\section{Acknowledments}
The author was partially supported by Fondecyt Grant \# 1141221.
Useful conversations with Alberto Faraggi and Stefan Theisen
are gratefully acknowledge. The author would also like to thank Macarena Lagos for a careful reading of the manuscript.   

\appendix

\section{The equations. }
\label{equations}

Using the notation
\begin{equation}
{\partial^2 V(q_i) \over \partial q_i \partial q_j } = V_{ij} 
\end{equation} 
where $q_1,q_2,q_3$ are the invariants of $Y=\eta^{-1}g$ the 10 equations following from (\ref{hessq}) are:
\begin{eqnarray}
{\cal E}_{1} &=&  
-4V_{11}V_{22}V_{1}q_{1}q_{3}-4V_{11}V_{22}q_{2}q_{3}V_{2}+8V_{11}V_{23}V_{1}q_{2}q_{3}-8V_{12}V_{1}q_{2}V_{13}q_{3}-V_{1}^3\nonumber\\&&
-8V_{12}^2q_{3}^3V_{33}-4V_{12}^2V_{1}q_{2}^2-4f_{12}^2q_{3}^2V_{3}-4V_{12}V_{1}^2q_{2}-4V_{1}^2V_{13}q_{3}-2q_{3}V_{22}V_{1}^2- \nonumber\\&&
8q_{3}^3V_{22}V_{13}^2-4V_{1}V_{13}^2q_{3}^2-8V_{11}q_{3}^3V_{23}^2-2V_{11}q_{1}V_{1}^2-2V_{11}q_{3}V_{2}^2+8V_{11}V_{22}q_{3}^3V_{33}+\nonumber\\&&
4V_{11}V_{22}V_{1}q_{2}^2+ 4V_{11}V_{22}q_{3}^2V_{3}-8V_{11}V_{23}q_{3}^2V_{2}+4V_{11}V_{1}V_{33}q_{3}^2+2V_{11}V_{1}q_{2}V_{2}+\nonumber\\&&
 2V_{11}V_{1}V_{3}q_{3}+4V_{12}^2V_{1}q_{1}q_{3}+4V_{12}^2q_{2}q_{3}V_{2}+8V_{12}q_{3}^2V_{23}V_{1}+16V_{12}q_{3}^3V_{23}V_{13}+\nonumber\\&&
4V_{12}q_{3}V_{2}V_{1}+8V_{12}q_{3}^2V_{2}V_{13}-8q_{3}^2V_{22}V_{1}V_{13}     \nonumber\\
\
{\cal E}_{2} &=&  4V_{1}^2V_{12}-V_{1}^2V_{3}-V_{1}V_{2}^2-8V_{1}V_{12}q_{3}q_{1}V_{23}-8V_{1}V_{12}q_{2}V_{33}q_{3}\nonumber\\&&
-4V_{1}V_{11}q_{1}V_{33}q_{3}+8V_{1}q_{3}V_{22}q_{1}V_{13}+8V_{1}q_{3}q_{2}V_{23}V_{13}-8q_{3}V_{12}V_{2}q_{1}V_{13}+\nonumber\\&&
8q_{3}V_{12}V_{2}V_{23}q_{2}-32q_{3}^2V_{12}V_{23}V_{13}q_{1}+16q_{3}V_{12}V_{23}V_{13}q_{2}^2+8V_{2}V_{11}q_{3}q_{1}V_{23}+\nonumber\\&&
4V_{2}V_{11}q_{2}V_{33}q_{3}-8V_{2}V_{13}q_{3}q_{2}V_{22}-16q_{1}q_{3}^2V_{11}V_{22}V_{33}-8q_{1}q_{3}V_{11}V_{22}V_{3}+\nonumber\\&&
8q_{2}^2V_{11}V_{22}V_{33}q_{3}-4V_{1}V_{12}^2q_{1}^2+8V_{1}V_{12}^2q_{2}-2V_{1}V_{2}V_{11}+4V_{1}q_{3}^2V_{23}^2-4V_{12}^2q_{3}V_{2}\nonumber\\&&
-4V_{12}^2q_{2}^2V_{3}-2V_{1}^2V_{33}q_{3}-4V_{2}^2V_{13}q_{3}+2V_{1}^2q_{1}V_{22}-2V_{1}V_{11}q_{1}V_{3}-4V_{1}V_{22}V_{33}q_{3}^2-\nonumber\\&&
2V_{1}q_{3}V_{22}V_{3}+4V_{1}q_{3}q_{1}V_{13}^2+16V_{12}^2q_{1}q_{3}^2V_{33}+8V_{12}^2q_{1}q_{3}V_{3}-8V_{12}^2q_{2}^2V_{33}q_{3}+\nonumber\\&&
8q_{3}^2V_{12}V_{2}V_{33}+4q_{3}V_{12}V_{2}V_{3}+4V_{2}V_{11}q_{3}V_{22}+2V_{2}V_{11}q_{2}V_{3}-8V_{2}V_{13}q_{3}^2V_{23}-\nonumber\\&&
4V_{2}V_{13}^2q_{3}q_{2}+16q_{1}q_{3}^2V_{11}V_{23}^2+16q_{1}q_{3}^2V_{13}^2V_{22}+4q_{2}^2V_{11}V_{22}V_{3}-8q_{2}^2V_{11}q_{3}V_{23}^2\nonumber\\&&-8q_{2}^2q_{3}V_{13}^2V_{22}-4V_{1}V_{12}q_{1}V_{2}+8V_{1}V_{12}V_{13}q_{3}-4V_{1}V_{12}q_{2}V_{3}-2V_{1}V_{2}q_{2}V_{22}-\nonumber\\&&
8V_{1}V_{11}q_{2}V_{22}+4V_{1}V_{11}V_{22}q_{1}^2-8V_{1}V_{11}q_{3}V_{23}    \nonumber\\
\
{\cal E}_{3} &=& -V_{2}^2V_{3}-4V_{12}^2V_{1}-4V_{2}V_{22}q_{2}V_{33}q_{3}-8V_{2}V_{12}q_{1}V_{33}q_{3}+8V_{2}V_{23}V_{13}q_{1}q_{3}+\nonumber\\&&
8V_{22}V_{33}q_{3}V_{11}q_{1}^2-16V_{22}V_{11}q_{2}V_{33}q_{3}+4V_{22}V_{33}q_{3}V_{1}q_{1}+16V_{12}q_{3}V_{23}V_{13}q_{1}^2-\nonumber\\&&
32V_{12}q_{3}q_{2}V_{23}V_{13}+2V_{2}V_{22}V_{1}+8V_{12}^2q_{2}V_{3}-2V_{2}V_{3}V_{11}-2V_{2}^2V_{33}q_{3}-4V_{12}^2V_{3}q_{1}^2+\nonumber\\&&
4V_{22}V_{1}V_{11}+4V_{2}q_{3}V_{13}^2+4V_{12}V_{1}V_{3}-2V_{2}V_{22}q_{2}V_{3}+8V_{2}V_{22}V_{13}q_{3}-4V_{2}V_{12}q_{1}V_{3}-\nonumber\\&&
8V_{2}V_{12}q_{3}V_{23}-4V_{2}V_{11}V_{33}q_{3}+4V_{2}q_{3}q_{2}V_{23}^2-8V_{22}V_{11}q_{2}V_{3}+4V_{22}V_{3}V_{11}q_{1}^2+\nonumber\\&&
2V_{22}V_{3}V_{1}q_{1}-8V_{22}q_{3}V_{13}^2q_{1}^2+16V_{22}V_{13}^2q_{3}q_{2}-8V_{12}^2V_{33}q_{3}q_{1}^2+16V_{12}^2q_{2}V_{33}q_{3}+\nonumber\\&&
8V_{12}V_{1}V_{33}q_{3}-8q_{3}V_{23}^2V_{11}q_{1}^2+16q_{3}V_{23}^2V_{11}q_{2}-4q_{3}V_{23}^2V_{1}q_{1}-8q_{3}V_{23}V_{1}V_{13}      \nonumber\\
\
{\cal E}_{4} &=&  4V_{11}V_{22}V_{33}q_{3}+2V_{11}V_{22}V_{3}+2V_{22}V_{2}V_{33}q_{3}+V_{22}V_{2}V_{3}-4q_{3}V_{13}^2V_{22}-\nonumber\\&&
4V_{11}q_{3}V_{23}^2-4V_{12}^2V_{33}q_{3}-2V_{12}^2V_{3}+8q_{3}V_{12}V_{13}V_{23}-2V_{2}q_{3}V_{23}^2     \nonumber\\
\
{\cal E}_{5} &=&  -V_{1}^2V_{2}+V_{1}^2V_{11}-2V_{1}V_{11}V_{22}q_{1}q_{2}-4V_{1}V_{11}q_{3}q_{1}V_{23}+4V_{1}V_{12}q_{3}q_{1}V_{13}+\nonumber\\&&
4q_{2}V_{11}V_{2}q_{3}V_{23}-4q_{2}V_{11}V_{22}V_{33}q_{3}^2-2q_{2}V_{11}q_{3}V_{22}V_{3}-8q_{2}V_{12}V_{13}q_{3}^2V_{23}-\nonumber\\&&
4q_{2}V_{12}V_{13}q_{3}V_{2}-2V_{1}V_{12}^2q_{3}+q_{2}V_{11}V_{2}^2-2V_{12}^2q_{2}^2V_{2}-V_{1}^2q_{2}V_{22}-2V_{1}^2q_{3}V_{23}+\nonumber\\&&
4q_{2}V_{12}^2V_{33}q_{3}^2+2q_{2}V_{12}^2V_{3}q_{3}+4q_{2}q_{3}^2V_{13}^2V_{22}-2V_{1}V_{11}q_{1}V_{2}+2V_{1}V_{11}q_{3}V_{22}+\nonumber\\&&
2V_{1}V_{12}^2q_{1}q_{2}-2V_{1}V_{12}q_{2}V_{2}+4V_{1}V_{12}V_{33}q_{3}^2+2V_{1}V_{12}V_{3}q_{3}-4V_{1}V_{13}q_{3}^2V_{23}-\nonumber\\&&
2V_{1}V_{13}q_{3}V_{2}+2V_{11}V_{2}q_{2}^2V_{22}+4q_{2}V_{11}q_{3}^2V_{23}^2     \nonumber\\
\
{\cal E}_{6} &=&   -8V_{22}V_{11}V_{33}q_{3}q_{1}q_{2}-16V_{12}q_{3}V_{23}V_{13}q_{1}q_{2}-2V_{11}V_{2}^2+8V_{2}q_{3}V_{22}q_{1}V_{13}-\nonumber\\&&
4V_{2}V_{11}q_{1}V_{33}q_{3}-8V_{2}V_{12}q_{3}q_{1}V_{23}-4V_{22}V_{11}V_{3}q_{1}q_{2}-4V_{22}V_{1}q_{2}V_{33}q_{3}+\nonumber\\&&
8V_{22}q_{3}V_{13}^2q_{1}q_{2}+8V_{11}q_{3}V_{23}^2q_{1}q_{2}+8V_{12}^2V_{33}q_{3}q_{1}q_{2}-V_{2}^3+2V_{11}V_{1}V_{3}+\nonumber\\&&
4V_{12}^2V_{1}q_{1}-4q_{3}V_{1}V_{13}^2-24V_{11}q_{3}^2V_{23}^2-24V_{12}^2V_{33}q_{3}^2-12V_{12}^2V_{3}q_{3}+8V_{12}^2q_{2}V_{2}-\nonumber\\&&
V_{2}V_{1}V_{3}-4V_{2}^2q_{3}V_{23}-4V_{2}^2q_{1}V_{12}-2V_{2}^2q_{2}V_{22}-4V_{2}q_{3}^2V_{23}^2-4V_{2}V_{12}^2q_{1}^2+8V_{2}V_{12}V_{1}-\nonumber\\&&
24q_{3}^2V_{13}^2V_{22}+4V_{2}V_{11}V_{22}q_{1}^2+2V_{2}V_{22}V_{1}q_{1}+2V_{2}q_{3}V_{22}V_{3}+4V_{2}V_{22}V_{33}q_{3}^2-\nonumber\\&&
2V_{2}V_{11}q_{1}V_{3}-2V_{2}V_{1}V_{33}q_{3}+4V_{2}q_{3}q_{1}V_{13}^2-4V_{22}V_{11}V_{1}q_{1}-8V_{22}V_{1}V_{13}q_{3}-\nonumber\\&&
2V_{22}V_{1}q_{2}V_{3}+4V_{11}V_{1}V_{33}q_{3}+4V_{12}^2V_{3}q_{1}q_{2}+8V_{12}q_{3}V_{23}V_{1}+4q_{3}V_{1}q_{2}V_{23}^2-\nonumber\\&&
16V_{11}V_{2}q_{3}V_{23}-8V_{11}V_{2}q_{2}V_{22}+24V_{11}V_{22}V_{33}q_{3}^2+12V_{11}q_{3}V_{22}V_{3}+48V_{12}V_{13}q_{3}^2V_{23}+\nonumber\\&&
16V_{12}V_{13}q_{3}V_{2}    \nonumber\\
\
{\cal E}_{7} &=&  -2V_{12}^2V_{2}+2V_{12}^2q_{1}V_{3}+4V_{12}^2q_{1}V_{33}q_{3}+4V_{12}V_{2}V_{33}q_{3}+2V_{12}V_{2}V_{3}-\nonumber\\&&
8V_{12}V_{23}V_{13}q_{1}q_{3}+V_{22}V_{2}^2+2V_{2}V_{22}V_{11}-4V_{2}q_{3}V_{23}V_{13}-4q_{1}V_{11}V_{22}V_{33}q_{3}-\nonumber\\&&
2q_{1}V_{11}V_{22}V_{3}+4q_{1}V_{11}q_{3}V_{23}^2+4q_{1}q_{3}V_{13}^2V_{22}     \nonumber\\
\
{\cal E}_{8} &=&  4V_{1}V_{2}V_{11}+4V_{1}V_{11}q_{2}V_{22}+8V_{1}V_{11}q_{3}V_{23}-V_{1}V_{2}^2-4V_{1}V_{2}q_{3}V_{23}-\nonumber\\&&
2V_{1}V_{2}q_{2}V_{22}+2V_{1}q_{3}V_{22}V_{3}+4V_{1}V_{22}V_{33}q_{3}^2-4V_{1}V_{12}^2q_{2}-8V_{1}V_{12}V_{13}q_{3}-\nonumber\\&&
4V_{1}q_{3}^2V_{23}^2-2q_{1}V_{11}V_{2}^2-8V_{2}V_{11}q_{3}q_{1}V_{23}-4q_{1}V_{11}V_{2}q_{2}V_{22}+8q_{1}q_{3}^2V_{11}V_{22}V_{33}+\nonumber\\&&
4q_{1}q_{3}V_{11}V_{22}V_{3}-8q_{1}q_{3}^2V_{11}V_{23}^2-8V_{12}^2q_{1}q_{3}^2V_{33}-4V_{12}^2q_{1}q_{3}V_{3}+\nonumber\\&&
4q_{1}V_{12}^2q_{2}V_{2}+16q_{3}^2V_{12}V_{23}V_{13}q_{1}+8q_{3}V_{12}V_{2}q_{1}V_{13}-8q_{1}q_{3}^2V_{13}^2V_{22}     \nonumber\\
\
{\cal E}_{9} &=&  2V_{12}V_{2}^2-2V_{2}V_{11}q_{1}V_{22}+V_{2}V_{3}V_{11}+2V_{2}V_{11}V_{33}q_{3}+2V_{2}q_{1}V_{12}^2+\nonumber\\&&
4V_{2}V_{12}q_{3}V_{23}-4V_{2}V_{22}V_{13}q_{3}-2V_{2}q_{3}V_{13}^2-2V_{12}^2q_{2}V_{3}-4V_{12}^2q_{2}V_{33}q_{3}+\nonumber\\&&
8V_{12}q_{3}q_{2}V_{23}V_{13}+4V_{22}V_{11}q_{2}V_{33}q_{3}+2V_{22}V_{11}q_{2}V_{3}-4q_{3}V_{23}^2V_{11}q_{2}-4V_{22}V_{13}^2q_{3}q_{2}     \nonumber\\
\
{\cal E}_{10} &=&       V_{11}V_{2}^2+2V_{11}V_{2}q_{2}V_{22}+4V_{11}V_{2}q_{3}V_{23}-2V_{11}q_{3}V_{22}V_{3}-\nonumber\\&&
4V_{11}V_{22}V_{33}q_{3}^2+4V_{11}q_{3}^2V_{23}^2-2V_{12}^2q_{2}V_{2}+2V_{12}^2V_{3}q_{3}+4V_{12}^2V_{33}q_{3}^2-\nonumber\\&&
8V_{12}V_{13}q_{3}^2V_{23}-4V_{12}V_{13}q_{3}V_{2}+4q_{3}^2V_{13}^2V_{22}
\label{wq}
\end{eqnarray} 

The corresponding equations in the $k_i$-system are straightforward to compute but even longer, 
so we do not include them here.

\section{Quadratic combinations}
\label{Q}

The original equations arising from (\ref{hessq}) are cubic in derivatives of $V$. By taking appropriated linear combinations (with non-zero functional dependent coefficients) we are able to produce 6 quadratic equations and then 2 linear ones. The linear ones are specially useful to study the space of solutions since general theorems are available. 

The following set of quadratic equations can be deduced from (\ref{wq}):

In the $q_i$-system oV coordinates these equations are:
\begin{eqnarray}
{\cal Q}_1 &=& 2q_{3}^3V_{22}V_{33}-2q_{3}^3V_{23}^2+q_{3}^2V_{22}V_{3}-4q_{3}^2V_{13}V_{12}+4q_{3}^2V_{11}V_{23}+\nonumber\\&&
q_{3}V_{2}V_{11}+2q_{3}V_{11}q_{2}V_{22}-2q_{3}V_{12}^2q_{2}     \nonumber\\
{\cal Q}_2 &=&  4q_{3}^3V_{12}V_{33}-4q_{3}^3V_{13}V_{23}-2q_{3}^2V_{12}^2+4q_{3}^2V_{12}V_{13}q_{1}+2q_{3}^2V_{12}V_{3}-\nonumber\\&&
4q_{3}^2V_{11}q_{1}V_{23}+2q_{3}^2V_{22}V_{11}+q_{3}^2V_{22}V_{2}-q_{3}q_{1}V_{2}V_{11}-2q_{3}q_{1}V_{11}q_{2}V_{22}+2q_{3}V_{12}^2q_{1}q_{2}    \nonumber\\
{\cal Q}_3 &=&   2q_{3}^3V_{11}V_{33}+4q_{3}^3V_{23}V_{12}-2q_{3}^3V_{13}^2-4q_{3}^3V_{13}V_{22}-2q_{3}^2V_{11}q_{1}V_{22}+\nonumber\\&&
q_{3}^2V_{11}V_{3}+4q_{3}^2V_{11}q_{2}V_{23}+2q_{3}^2q_{1}V_{12}^2-4q_{3}^2V_{12}q_{2}V_{13}+2q_{3}^2V_{12}V_{2}+\nonumber\\&&
q_{3}q_{2}V_{2}V_{11}+2q_{3}V_{11}q_{2}^2V_{22}-2q_{3}V_{12}^2q_{2}^2   \nonumber\\
{\cal Q}_4 &=&  4q_{3}^2V_{11}V_{33}+8q_{3}^2V_{23}V_{12}-4q_{3}^2V_{13}^2-8q_{3}^2V_{13}V_{22}-4q_{3}V_{11}q_{1}V_{22}+\nonumber\\&&
8q_{3}V_{11}q_{2}V_{23}+2q_{3}V_{11}V_{3}+4q_{3}q_{1}V_{12}^2+4q_{3}V_{12}V_{2}-8q_{3}V_{12}q_{2}V_{13}-4q_{3}V_{1}V_{13}-\nonumber\\&&
2q_{3}V_{1}V_{22}-2V_{11}q_{1}V_{1}+4V_{11}q_{2}^2V_{22}+2q_{2}V_{2}V_{11}-V_{1}^2-4V_{1}q_{2}V_{12}-4V_{12}^2q_{2}^2    \nonumber\\
{\cal Q}_5 &=&   4q_{3}^2V_{12}V_{33}-4q_{3}^2V_{13}V_{23}-4q_{3}V_{11}q_{1}V_{23}+2q_{3}V_{22}V_{11}-2q_{3}V_{12}^2+\nonumber\\&&
4q_{3}V_{12}V_{13}q_{1}+2q_{3}V_{12}V_{3}-2q_{3}V_{2}V_{13}-2q_{3}V_{1}V_{23}+2V_{12}^2q_{1}q_{2}-2q_{2}V_{2}V_{12}-\nonumber\\&&
2q_{1}V_{11}q_{2}V_{22}-2q_{1}V_{2}V_{11}+V_{11}V_{1}-V_{1}q_{2}V_{22}-V_{1}V_{2}   \nonumber\\
{\cal Q}_6 &=&   -4q_{3}^2V_{23}^2+4q_{3}^2V_{22}V_{33}+8q_{3}V_{11}V_{23}+2q_{3}V_{22}V_{3}-8q_{3}V_{13}V_{12}-\nonumber\\&&
4q_{3}V_{2}V_{23}+4V_{11}q_{2}V_{22}-2q_{2}V_{22}V_{2}-4V_{12}^2q_{2}-V_{2}^2+4V_{2}V_{11} 
\label{Qq}
\end{eqnarray} 
Here $Vij = {\partial V(q) \over \partial q_i \partial q_j}$, etc.  

The corresponding equations in the k-system are straightforward to compute but even longer, 
so we do not include them here. 

\section{Linear combinations}
\label{L}

Finally, taking combinations of the quadratic equations we can produce two linear equations. 
In the $q_i$-system of coordinates these equations are:
\begin{eqnarray}
{\cal L}_1 &=&  -V_{1}-4V_{13}q_{3}-4q_{2}V_{12}-2q_{3}V_{22}-2q_{1}V_{11}   \label{L1a}\\
{\cal L}_2 &=&  -V_{2}+2V_{11}-4q_{3}V_{23}-2q_{2}V_{22}   \label{L2a}
\end{eqnarray} 
Here $Vij = {\partial V(k) \over \partial k_i \partial k_j}$, etc.  The simplicity of these equations should be remarked. Yet, their solutions are not easy, and most importantly, the do not form a complete set.

In the $k_i$-system of coordinates these equations are:
\begin{eqnarray}
{\cal L'}_1 &=&  -2k_{2}^2V_{12}+((-2V_{13}-2V_{22})k_{3}-V_{11}k_{1})k_{2}-2k_{3}^2V_{23}+V_{11}k_{3}      \nonumber\\
{\cal L'}_2 &=&      (-2V_{13}-2V_{23}k_{1}-V_{22})k_{3}-k_{2}(2V_{12}+V_{22}k_{1})
\label{Lk}
\end{eqnarray}


\end{document}